\documentclass[12pt]{iopart}
\usepackage{iopams}
\usepackage{hyperref}
\usepackage{graphicx}

\begin{document}
\title[Pattern formation in a two-dimensional two-species diffusion model]{Pattern formation in a two-dimensional two-species diffusion model with anisotropic nonlinear diffusivities: a lattice approach}

\author{Yuri~Yu~Tarasevich$^1$, Valeri~V~Laptev$^{1,2}$, Andrei~S~Burmistrov$^1$, Nikolai~I~Lebovka$^{3,4}$}
\address{$^1$Astrakhan State University, 20A Tatishchev Street, Astrakhan, 414056, Russia}
\address{$^2$Astrakhan State Technical University, 16 Tatishchev Street, Astrakhan, 414025, Russia}
\address{$^3$F. D. Ovcharenko Institute of Biocolloidal Chemistry, NAS of Ukraine, 42 Boulevard Vernadskogo, 03142 Kiev, Ukraine}
\address{$^4$Taras Shevchenko Kiev National University, Department of Physics, 64/13 Volodymyrska Street, 01601 Kiev, Ukraine}
\ead{tarasevich@asu.edu.ru}
\vspace{10pt}
\begin{indented}
\item[]June 2017
\end{indented}

\begin{abstract}
Diffusion in a two-species two-dimensional system has been simulated using a lattice approach. Rodlike particles were considered as linear $k$-mers of two mutually perpendicular orientations ($k_x$- and $k_y$-mers) on a square lattice. These $k_x$- and $k_y$-mers were treated as species of two kinds.
A random sequential adsorption model was used to produce an initial homogeneous distribution of $k$-mers. The concentration of $k$-mers, $p$, was varied in the range from 0.1 to the jamming concentration, $p_j$.
By means of the Monte Carlo technique, translational diffusion of the $k$-mers was simulated as a random walk, while rotational diffusion was ignored. We demonstrated that the diffusion coefficients are strongly anisotropic and nonlinearly concentration-dependent. For sufficiently large concentrations (packing densities) and $k \geq 6$, the system tends toward a well-organized steady state. Boundary conditions predetermine the final state of the system.
When periodic boundary conditions are applied along both directions of the square lattice, the system tends to a steady state in the form of diagonal stripes. The formation of stripe domains takes longer time the larger the lattice size, and is observed only for concentrations above a particular critical value.
When insulating (zero flux) boundary conditions are applied along both directions of the square lattice, each kind of $k$-mer tries to completely occupy a half of the lattice divided by a diagonal, e.g., $k_x$-mers locate in the upper left corner, while the $k_y$-mers are situated in the lower right corner (``yin-yang'' pattern). From time to time, regions built of $k_x$- and $k_y$-mers exchange their locations through irregular patterns.
When mixed boundary conditions are used (periodic boundary conditions are applied along one direction whereas insulating boundary conditions are applied along the other one), the system still tends to form the stripes, but they are unstable and change their spatial orientation.
\end{abstract}
\submitto{\jpa}
\section{Introduction}\label{sec:intro}

Since Turing's pioneering work regarding pattern formation~\cite{Turing1952}, numerous research papers have been devoted to two-species reaction-diffusion (RD) systems.  The Turing instability of a homogeneous steady state is one of the best understood theoretical mechanisms for pattern formation. Nonequilibrium pattern formation in chemical systems is based on the interplay between the reactions and diffusion (see, e.g., \cite{Mikhailov2009} for a review). In any set of RD partial differential equations, the reaction terms provide a nonlinear coupling between equations. Nevertheless, a chemical reaction is not the sole mechanism that can induce such nonlinear feedback~\cite{Andelman2009JPhChB}. For instance, cross-diffusion may be quite important in pattern formation~\cite{Vanag2009}. Pattern formation in  RD systems can be considered using a lattice approach~\cite{Odor2004RMP}.

Systems composed of shape-anisotropic (particularly, rodlike) particles demonstrate orientational order and self-organization~\cite{Boerzsoenyi2013}. A range of experimental studies for two-dimensional (2D) systems of vertically vibrated rodlike particles revealed different kind of rearrangement and pattern formation in such systems. For example, a horizontal monolayer of macroscopic elongated particles of various shapes with aspect ratios $k$ ranging from $4$ to $12.6$ have been experimentally tested~\cite{Narayan2006}. These experiments demonstrated that, at high packing fraction, the shape of the particles is important for the orientational ordering.

The orientational order in vertically agitated granular-rod monolayers has also been investigated  experimentally~\cite{Muller2015PRE}. The results have been compared quantitatively using both equilibrium Monte Carlo (MC) simulations and density functional theory. When the density is sufficiently high, short rods form tetratic arrangements, while long rods form uniaxial nematic state.  In both experiments and simulations, the length-to-width ratio at which the order changes from tetratic to uniaxial was about 7.3. The universality of this agreement has been illustrated for the ordering of rod-like particles across equilibrium and nonequilibrium systems. The assembly of granular rods into ordered states has been found to be independent of either  agitation frequency or strength. This  suggests that the specific nature of energy injection into such a nonequilibrium system does not play a crucial role~\cite{Muller2015PRE}.

Stainless steel rods with aspect ratios of $k=20, 40$, and $60$ confined to 2D containers  have been analysed~\cite{Galanis2006,Galanis2010}. At high packing densities, the distinct patterns reflected competition between bulk nematic and boundary alignments. In a container of circular geometry, this competition produced a bipolar configuration with two diametrically opposed point defects. As packing density increased, the patterning shifted from bipolar to a uniform alignment~\cite{Galanis2010}. The presence of large-scale collective swirl motions has been revealed in monolayers of vibrated granular rods (a range of rice species, mustard seeds and stainless steel rods with aspect ratios from $\approx 1$ to $\approx 8$)~\cite{Aranson2007}. The authors speculated that the very strong sensitivity of swirling to the shape of the particles could  be related to the formation of tetratic structures.

The self-diffusion of colloidal rodlike virus in an isotropic and nematic phase has been experimentally studied~\cite{Lettinga2005EPL}. The ratio $D_\parallel/ D_\perp$ increases monotonically with increasing virus concentration, where $D_\parallel$ and $D_\perp$ are the diffusivities parallel and perpendicular to the nematic director, respectively.
Similar results were obtained for granular rods on a substrate with experiments using a mono-layer of bead chains in a vibrated container~\cite{Yadav2012EPJE}. Numerous additional examples of patterns and of collective behaviour in granular media together with appropriate references can be found in the review~\cite{Aranson2006RMP}.

Thus, pattern formation is observed not only in RD systems but also in systems without any chemical reactions between the species. The question is which effect provides nonlinear feedback? Nonlinear cross-diffusion may be considered as a possible candidate. This suggestion is based on a number of experiments~\cite{Yadav2012EPJE}. When the diffusion of granular rod in two dimensions was studied using linked chains on a vibrated substrate, it was shown that aspect ratio and packing density have significant effects on the diffusivity~\cite{Yadav2012EPJE}.

In recent decades, much attention has been paid to the study of self-assembly in systems of linear $k$-mers (particles occupying $k$ adjacent adsorption sites) deposited on 2D lattices. A linear $k$-mer represents the simplest model of an elongated particle with an aspect ratio of $k$. Computer simulations have been extensively applied to investigate percolation and jamming phenomena for the random sequential absorption (RSA) of $k$-mers (see, e.g.,~\cite{Centres2015JStatMech,Kuriata2016,Budinski2017PRE} and the references therein).

Recently, diffusion-driven pattern formation in a 2D system of $k$-mers has been studied by means of MC simulation~\cite{Lebovka2017PRE}. Periodic boundary conditions (PBCs) were applied along both directions of a square lattice. Equal numbers of $k_x$-mers and $k_y$-mers were assumed. The concentration of $k$-mers was the maximal obtainable for an RSA mechanism, i.e., systems at jammed states were considered. Notice, that other mechanisms, e.g., RSA with diffusional relaxation can produce more dense systems~\cite{Fusco2001JChPh}. Pattern formation as stripe domains was observed only for $k \geq 6$.  Nevertheless, some important questions still remain:
\begin{enumerate}
\item Why is $k = 6$ the critical length? Which characteristics change critically when $k$ is changed from 5 to 6? (As a potentially interesting characteristic, the diffusion coefficient is particularly prominent. Its anisotropy can have an effect on self-organization.)
\item  What is the effect of concentration on such pattern formation? Is self-organization possible in fairly dilute systems?
\item Is the finite-size effect crucial?  It is not yet completely clear, whether patterns form for $L \gg k$. Although there are clear signs of reorganization, there is no reliable evidence that patterns arise as a result of any particular initial conditions. What happens in the continuous limit $k / L \to 0$? Does the increase in the size of the system only lead to an increase in the relaxation time, or do qualitative changes in its behaviour arise?
    \item What happens when the number of $k_x$-mers is not equal to the number of $k_y$-mers?
\item What is the effect of topology on the pattern formation? What happens when the boundary conditions change? (PBCs are radically different from other types of boundary conditions, since they introduce translational symmetry into the system in two directions.)
\item Is it possible to construct a continuous analogue of the model that has similar behaviour to the lattice model?
\end{enumerate}

We can speculate that, for rodlike particles with two mutually perpendicular orientations on a  square lattice, pattern formation is the result of nonlinearity in the diffusion coefficients. In the present research, our main goal is to elicit whether random walks provide nonlinear feedback and produce self-organization. Additionally, this paper analyses how different kinds of boundary conditions affect the pattern formation in a two-species diffusion system. We suggest that pattern formation is a result of correlation of the movements of the $k$-mers. These correlations can occur only in fairly dense systems of moderate size in the presence of translational symmetry (PBCs). For this reason, we examine the effect of concentration (packing density) and lattice size on the pattern formation.

In our study, a lattice approach is used. The diffusion of rodlike particles is simulated by means of kinetic MC simulation. The rest of the paper is constructed as follows. In section~\ref{sec:methods}, the technical details of the simulations are described and all necessary quantities are introduced. Section~\ref{sec:results} presents our principal findings. Section~\ref{sec:conclusion} summarizes the main results.

\section{Details of simulation}\label{sec:methods}

\subsection{Lattice model}\label{subsec:common}

The problem was approached using a square lattice of size $L\times L$. Most calculations were performed using $L=256$. For one particular value of $k$ ($k=12$), we examined the scaling effect using $L=256, 512, 1024,$ and $2048$. We used a range of boundary conditions:
\begin{enumerate}
  \item toroidal boundary conditions, i.e., periodic boundary conditions along both the $x$ and $y$ axes (PBCs),
  \item insulating (zero flux) boundary conditions along both the $x$ and $y$ axes (IBCs),
  \item mixed boundary conditions, i.e., periodic boundary conditions  along the $x$ axis and insulating boundary conditions along the $y$ axis (MBCs).
\end{enumerate}

The RSA model~\cite{Evans1993RMP} was used to produce an initial homogeneous distribution of linear $k$-mers (i.e., rodlike particles occupying $k$ adjacent sites) in a 2D film. The rodlike particles were deposited randomly and sequentially, and their overlapping with previously placed particles was forbidden. In most experiments, the concentration of the particles corresponded to the jamming state, $p_j$. In this state, no additional $k$-mer can be placed because the presented voids are too small or of inappropriate shape. We additionally studied the effect of concentration on self-organization using the concentrations $p \in [0.1, p_j]$.

The length of the $k$-mers (aspect ratio)  was varied from $2$ to $12$. Generally, isotropic orientation of the  $k$-mers was assumed, i.e., that $k$-mers oriented along the $x$ and $y$ directions ($k_x$-mers and $k_y$-mers, respectively) were equiprobable in their  deposition. This corresponded to the zero value of a mean order parameter of the system, defined as
\begin{equation}\label{eq:s}
s = \frac{N_y - N_x}{N},
\end{equation}
where $N_x$ and $N_y$  are the numbers of $k_x$-mers and $k_y$-mers, respectively, and $N=N_y + N_x$ is the total number of $k$-mers. We also investigated the effect of anisotropy on pattern formation.

The diffusion of $k$-mers was simulated using the kinetic MC procedure. In our simulation, only translational diffusion was taken into consideration. This is essentially the case for fairly dense systems in the jamming state, where rotational diffusion is impeded, especially for large values of~$k$. Undoubtedly, for dilute systems, rotational diffusion can occur, however, this was ignored in our study.

Since our goal is not only to obtain the static equilibrium properties but also the dynamic correlation functions of the model, we did not consider every $k$-mer at each MC step~\cite[pp.~71,72]{LandauBinder}. By contrast, an arbitrary $k$-mer was randomly chosen at each step and a translational shift by one lattice unit along either the longitudinal ($\parallel$) or the transverse ($\perp$) axis of the $k$-mer was attempted. Equal probabilities to choose any of all the four possible directions to shift the $k$-mer were assumed. One time step of the MC computation, which corresponds to an attempted displacement of the total number of $k$-mers in the system, $N$, was taken as the MC time unit. Time  counting was started from the value of $t_{MC}=0$, being the initial moment (before diffusion), and the total duration of the simulation was typically $10^7$ MC time units.

\subsection{Determination of diffusivities}\label{subsec:diff}
To determinate the components of the diffusivity tensor, Einstein--Smoluchovski formula has been applied
\begin{equation}\label{eq:ES}
  \langle \Delta i^2 \rangle = 2 D_i t,
\end{equation}
where $i = x,y$ corresponds to the two possible directions, $t$ is the time, and $D_i$ is the diffusion coefficient (diffusivity) in the $x$ or $y$ direction, respectively. The left-hand side of \eref{eq:ES} is the mean square displacement (MSD) of a particle from its original position. Due to the finite-size of the system under consideration, the total displacement of the particle cannot exceed $L/2$ along a direction with PBCs and $L$ along a direction with IBCs.

To elicit the coefficients of diffusion, we considered strongly anisotropic systems, $s=1$. To compute the components of the coefficient of self-diffusion, $D^s_i$, the lattice was filled with $k_x$-mers until a given concentration. After that, the displacements of all the $k_x$-mers were  monitored at each given MC step. The MSD is a result of averaging over $10^2$ independent runs and all $k_x$-mers. $\langle \Delta x^2 \rangle = 2 D^s_x t_{MC}$ and $ \langle \Delta y^2 \rangle = 2 D^s_y t_{MC}$ have been fitted by a linear function using the least-square method.

To compute the components of the coefficient of cross-diffusion, $D^c_i$, a single $k_y$-mer (tracer) has been deposited onto the lattice and, after that, the lattice has been filled with $k_x$-mers until a given  concentration. Then, at each given MC step, the displacement of the tracer has been determined. The MSD is a result of averaging over $10^4$ independent runs.

\subsection{Quantities under consideration}\label{subsec:quantities}

For characterization of the rearrangement of the system under consideration, several quantities were monitored at each given MC step:
\begin{enumerate}
\item  The number of clusters, $n$, built of the same kinds of $k$-mers, i.e., clusters of $k_x$-mers and clusters of $k_y$-mers. We used the Hoshen--Kopelman algorithm~\cite{HK76} to find the cluster distribution. To facilitate comparison of the curves for the different values of $k$, we used normalized numbers of clusters, i.e., the current value divided by the value at the initial state, $t_{MC} = 0$.
    \item The local anisotropy, $s$, i.e., the order parameter $s$ \eref{eq:s} calculated in a window of $l \times l$ sites and averaged over the entirely set of windows.
   \item The fraction of interspecific contacts $n_{xy}^{\ast} = n_{xy} / ( n_{xy} + n_x + n_y )$, where $n_{xy}$ is the number of interspecific contacts between the different sorts of $k$-mers (i.e., $k_x$--$k_y$), $n_x$ and $n_y$  are the numbers of intraspecific contacts between $k$-mers of the same kind (i.e., $k_x$--$k_x$ and $k_y$--$k_y$, respectively).
  \item The shift ratio, $R$, i.e., the ratio of the number of shifts of the $k$-mers along the transverse axes to the number of shifts along their longitudinal axes during one MC step.
  \item The electrical conductivities, $\sigma$. We used the model described in~\cite{Lebovka2017PRE} to transform the system under consideration into a random resistor network (RRN). The Frank--Lobb algorithm~\cite{Frank1988PRB} was applied to calculate the electrical conductivity of such RRNs.
\end{enumerate}
All the quantities under consideration were averaged over 100 independent statistical runs, unless otherwise explicitly specified in the text.

\subsection{Estimation of the finite-size effect}\label{subsec:scaling}

\begin{figure}[htbp]
  \centering
  \includegraphics[width=\linewidth]{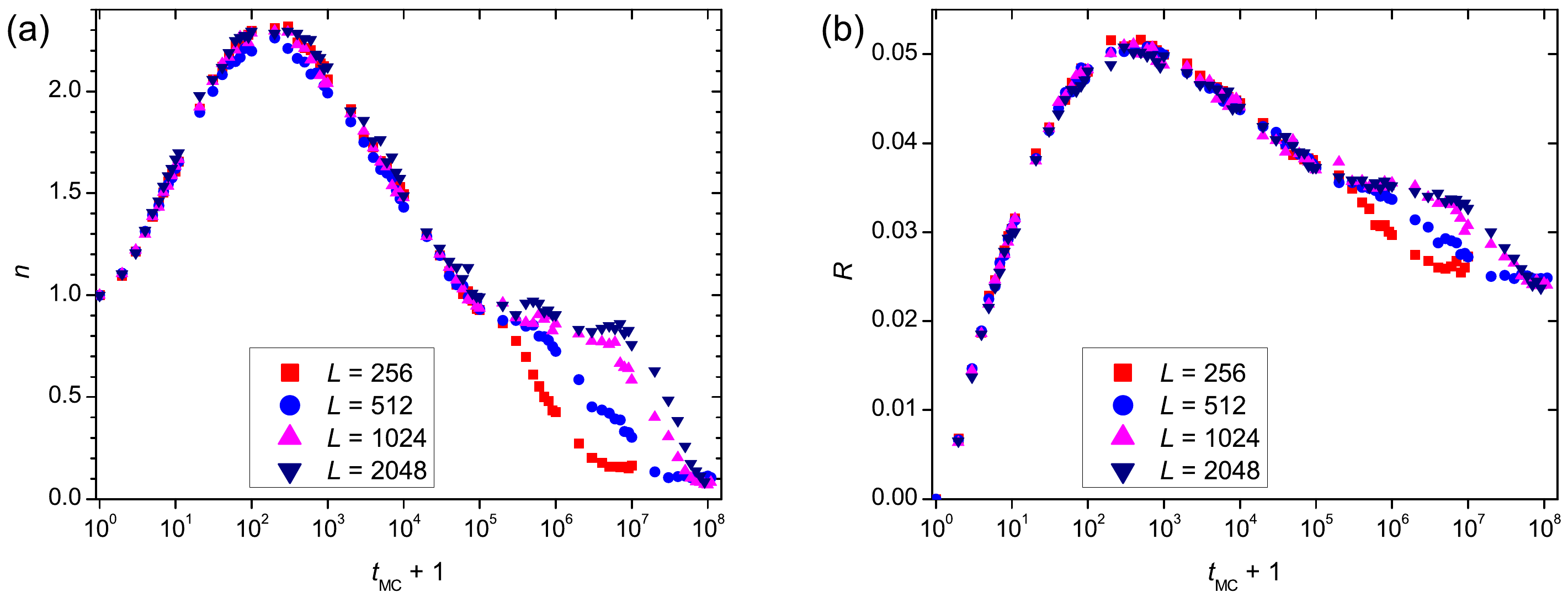}
  \caption{Normalized number of clusters, $n$, and shift ration, $R$, vs MC steps, $t_{MC}$, for different lattice sizes when PBCs are applied. $k=12$, $L=256$ (100 independent runs), 512 (25 independent runs), 1024 (5 independent runs), and 2048 (one run).}\label{fig:latticesizes}
\end{figure}

To determine if there were any finite-size effect, we performed computations for different lattices sizes, $L=256, 512, 1024,$ and $2048$, and a fixed value of $k$, $k=12$, when PBCs were applied. We chose the largest value of $k$ because the finite-size effect was expected to be more pronounced for this case. The results presented in figure~\ref{fig:latticesizes} suggest that the finite-size effect is important only during the last stage of the transient mode,  $10^5 \lesssim t_{MC} \lesssim 10^8$.

Since simulation using large lattices such as $L=2048$ is quite time-consuming, this makes it difficult to obtain sufficient statistical data, we basically used lattices of moderate size $L=256$ to collect the statistics for the principal investigations.

\section{Results}\label{sec:results}
\subsection{Diffusion coefficients}\label{subsec:nladiff}
\begin{figure}[htbp]
  \centering
  \includegraphics[width=\linewidth]{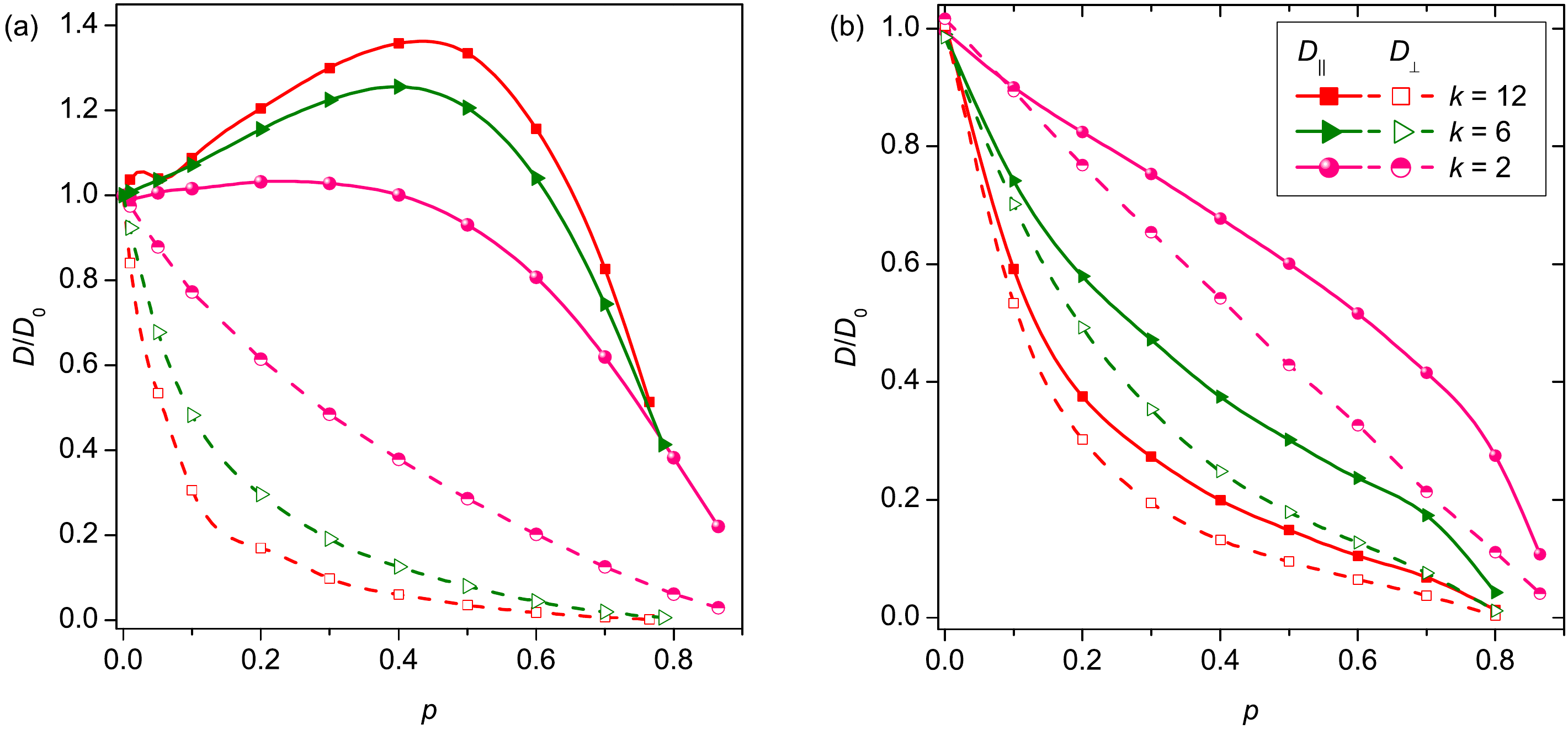}\\
  \caption{Examples of the dependencies of the coefficients of diffusion on concentration, $p$. $D_0$ is the coefficient of diffusion of the monomer. $L=256$, PBCs are applied. (a)~self-diffusion, (b)~cross-diffusion. The legend corresponds to both subfigures.  For clarity view, the results for intermediate values of $k$ have been omitted.}\label{fig:DiffCoeff}
\end{figure}

The simulations gave clear evidence that self-diffusion is strongly  anisotropic.
Both for self-diffusion and for cross-diffusion, the diffusion coefficients are strongly nonlinear concentration-dependent (figure~\ref{fig:DiffCoeff}). The transversal coefficients of self-diffusion, $D^s_\perp$, progressively decreased with increasing concentration, whereas the longitudinal ones, $D^s_\parallel$, went through a maximum  at $p \approx 0.45$, where $D^s_\parallel$ and $D^c_\parallel$ are the coefficients of diffusion when the particle moves along its axis (longitudinal shifts), $D^s_\perp$ and $D^c_\perp$  are the coefficients of diffusion when the particle moves perpendicular to its axis (transversal shifts). Thereby, the system under consideration is nonlinear and the diffusion coefficients are anisotropic.

\subsection{Periodic boundary conditions}\label{subsec:pbc}
The results presented in this section have been obtained for $L=256$, $p=p_j$, $s=0$, unless otherwise explicitly specified in the text. For all values of $k$, a transient mode and a steady state were observed.
\paragraph{Transition mode.} During the transient mode, the initial homogeneous jammed state transforms into a new state.

For $k \geq 6$, the transient mode is clearly divided into several stages of diffusive reorganization of the system under consideration.  We can identify the following  stages.
    \begin{description}
    \item[Fluidization.] During the first stage, the system undergoes drastic changes, namely, the jammed state (figure~\ref{fig:patternsk10}a) turns into an unjammed state.  Some $k$-mers play the role of capstones, i.e., when such a $k$-mer shifts from its initial position, the whole system of $k$-mers becomes more mobile. It is noteworthy that the  initial jammed state (figure~\ref{fig:patternsk10}a) and unjammed state (figure~\ref{fig:patternsk10}b) are visually almost indistinguishable, nevertheless, a number of clusters, $n$, (figure~\ref{fig:pbcallinone}(a)) clearly evidences that these states have very different connectivity. Upon the transition, the number of clusters, $n$, increases (figure~\ref{fig:pbcallinone}(a)), and the shift ratio, $R$, increases (figure~\ref{fig:pbcallinone}(d)). The end of this stage corresponds to the maxima of the curves in figure~\ref{fig:pbcallinone}(a) and figure~\ref{fig:pbcallinone}(d). Its duration is of the order of $100$ MC steps for large values of $k$ and of the order of 10 MC steps for small values of $k$. By contrast, the local anisotropy, $s$, does not change during this stage (figure~\ref{fig:pbcallinone}(c)).
\begin{figure}[htbp]
  \centering\small
  (a)\includegraphics[width=0.3\linewidth]{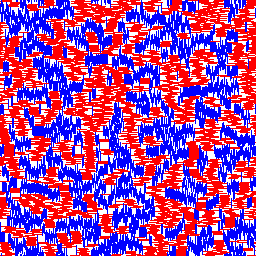}\hfill (b)\includegraphics[width=0.3\linewidth]{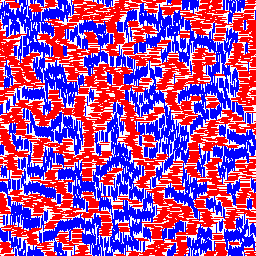}\hfill (c)\includegraphics[width=0.3\linewidth]{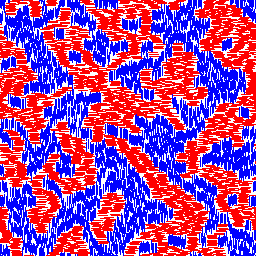}\\
  (d)\includegraphics[width=0.3\linewidth]{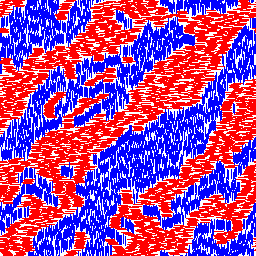}\hfill
  (e)\includegraphics[width=0.3\linewidth]{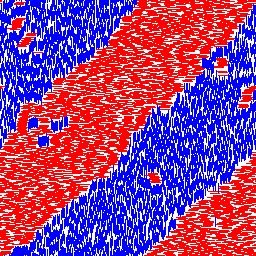}\hfill (f)\includegraphics[width=0.3\linewidth]{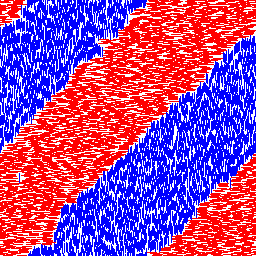}\\
  (g)\includegraphics[width=0.3\linewidth]{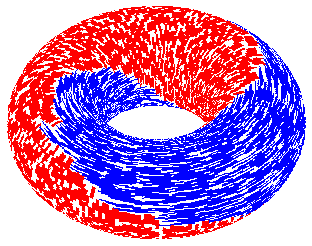}\\
  \caption{Temporal evolution of the system, $k=10$, $L=256$, PBCs. (a)~$t_{MC}=0$, initial jammed state, (b)~$t_{MC}=100$, end of fluidization, (c)~$t_{MC}=10^4$, labyrinth patterns, (d)~$t_{MC}=10^5$, labyrinth patterns are transforming into stripes, (e)~$t_{MC}=10^6$, final stage of stripe formation,  (f)~$t_{MC}=10^7$, steady state in the form of diagonal stripes, (g)~final ``yin-yang'' pattern on a torus (the same pattern as in (f)); for clarity, the major radius of the torus is significantly exaggerated.}\label{fig:patternsk10}
\end{figure}
\begin{figure}[htbp]
  \centering
  \includegraphics[width=\linewidth]{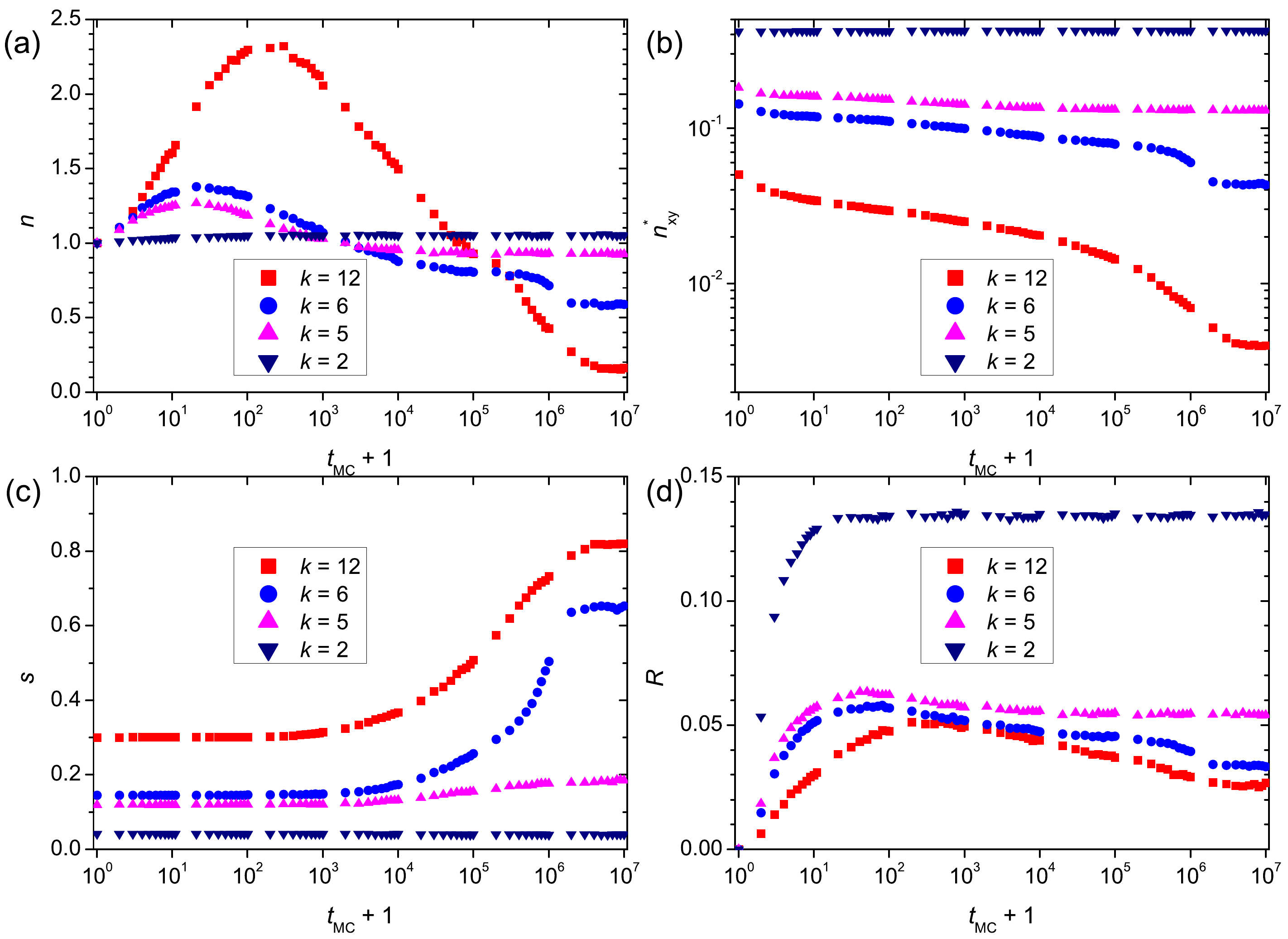}
  \caption{Examples of different quantities vs MC steps, $t_{MC}$, when PBCs are applied. (a)~Normalized number of clusters, $n$. (b)~Fraction of interspecific contacts, $n^\ast_{xy}$. (c)~Local anisotropy, $s$. (d)~Shift ratio, $R$.}\label{fig:pbcallinone}
\end{figure}
    \item[Coarsening.] During the second stage, the clusters coarsen and a labyrinthine structure forms  (figure~\ref{fig:patternsk10}c), the number of clusters, $n$, decreases (figure~\ref{fig:pbcallinone}a), the shift ratio, $R$, decreases  (figure~\ref{fig:pbcallinone}d), and the local anisotropy, $s$, increases (figure~\ref{fig:pbcallinone}c). The coarsening is complete after a time in the order of $10^5$ MC steps.
    \item[Mainland formation.] This stage is clearly recognizable only when the lattice size is large enough. During this stage, the labyrinth patterns transform into compact areas  with irregular borders (`mainlands'). For $L=256$, the narrow step in figure~\ref{fig:pbcallinone}a at $t_{MC} \sim 10^5$ corresponds to this stage. For larger values of $L$, this step is more easily distinguished, e.g., for $L=1024$, it lasts from $t_{MC}=10^5$ up to $t_{MC}=10^7$ (figure~\ref{fig:latticesizes}).
    \item[Pattern formation.] During the last stage, regular patterns begin to form. At the plane, these patterns appear as stripes. For $L=256$, the stripe formation is complete at a time in the order of $10^6$ MC steps. Its end corresponds to the transitions to horizontal of the curves in figures~\ref{fig:pbcallinone}a, \ref{fig:pbcallinone}d,
         \ref{fig:pbcallinone}c, and \ref{fig:pbcallinone}b.
    \end{description}
    The fraction of interspecific contacts, $n_{xy}^{\ast}$, decreases during the entirely transient mode (figure~\ref{fig:pbcallinone}b).  It is quite remarkable that the durations of the first two stages, i.e., fluidization and coarsening, are independent of the lattice size when the value of $k$ is fixed. By contrast, the mainland and stripe formation stages require more time the larger the size of the system (figure~\ref{fig:latticesizes}).
\paragraph{Steady state} During this stage, none of the quantities under consideration  demonstrate any tendency to increase or decrease but have essential fluctuations (final  horizontal parts of all the curves), the normalized number of clusters, $n$, and the fraction of interspecific contacts, $n^\ast_{xy}$, are very low, while the shift ratio, $R$, is stable (figures~\ref{fig:pbcallinone}a, \ref{fig:pbcallinone}d, \ref{fig:pbcallinone}c, and \ref{fig:pbcallinone}b). Local anisotropy is large, reflecting the occurrence of stripes, with a typical width of $L/2$, built of $k$-mers of the same orientation (figure~\ref{fig:patternsk10}f). The stripes become fairly stable, and on a plane, this pattern  corresponds to a torus divided into two equal parts (figure~\ref{fig:patternsk10}g).

For $k<6$, the number of observed stages is fewer than for the larger $k$-mers. Only the normalized number of clusters, $n$, and the shift ratio, $R$, allow clear identification of the  transient mode and the steady state (figures~\ref{fig:pbcallinone}(a) and (d)). The other quantities vary only insignificantly with time.

For $k=2$, the quantities $n$ and $R$ increase during $t_{MC} \lesssim 10$, after that no obvious  changes can be observed. For $k=3, 4,$ and $5$, the quantities $n$ and $R$ increase during $t_{MC} \lesssim 10$, but after that $n$ decreases during $t_{MC} \lesssim 100$ when $k=3,4$ and during $t_{MC} \lesssim 10^5$  when $k=5$.
For $k=5$, a smooth decreas of $n$ until  $\approx 0.9$ is observed. In contrast, for $k=2,3,$ and $4$, $n$ never goes below 1. Moreover, $n$ increases monotonically up to $1.05$ when $k=2$. In the final stage, $n$ is fixed for values of $k=2,3,4,$ and $5$, and this corresponds to a steady state.

\subsection{Effect of concentration}\label{subsec:concentration}
\begin{figure}[htbp]
  \centering\small
  (a)\includegraphics[width=0.3\linewidth]{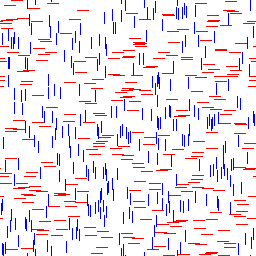}\hfill (b)\includegraphics[width=0.3\linewidth]{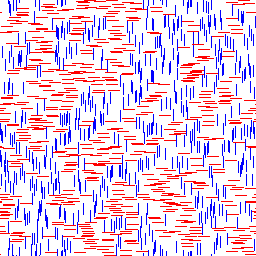}\hfill
  (c)\includegraphics[width=0.3\linewidth]{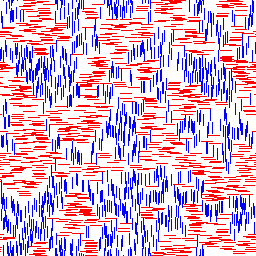}\\
  (d)\includegraphics[width=0.3\linewidth]{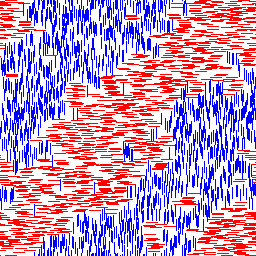}\hfill
  (e)\includegraphics[width=0.3\linewidth]{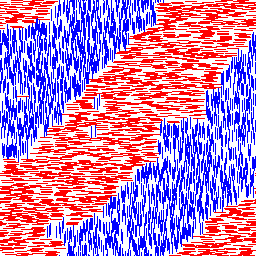}\hfill
  (f)\includegraphics[width=0.3\linewidth]{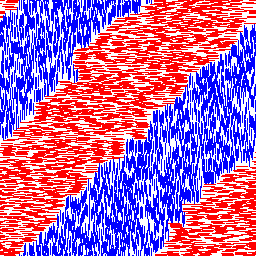}\\
    \caption{Examples of steady-state patterns for different concentrations, $p$. $k=12$, $L=256$, $t_{MC}=10^7$, PBCs. (a)~$p=0.1$, (b)~$p=0.2$, (c)~$p=0.3$, (d)~$p=0.4$, (e)~$p=0.5$, (f)~$p=0.6$.
}\label{fig:patternsconcentrations}
\end{figure}

We examined the effect of concentration on pattern formation.
Figure~\ref{fig:patternsconcentrations} presents examples of pattern formations  at different values of $p$ for $k=12$. We found that the stripes are observed only for fairly dense systems.

\begin{figure}[htbp]
  \centering\small
  (a)\includegraphics[width=0.45\linewidth]{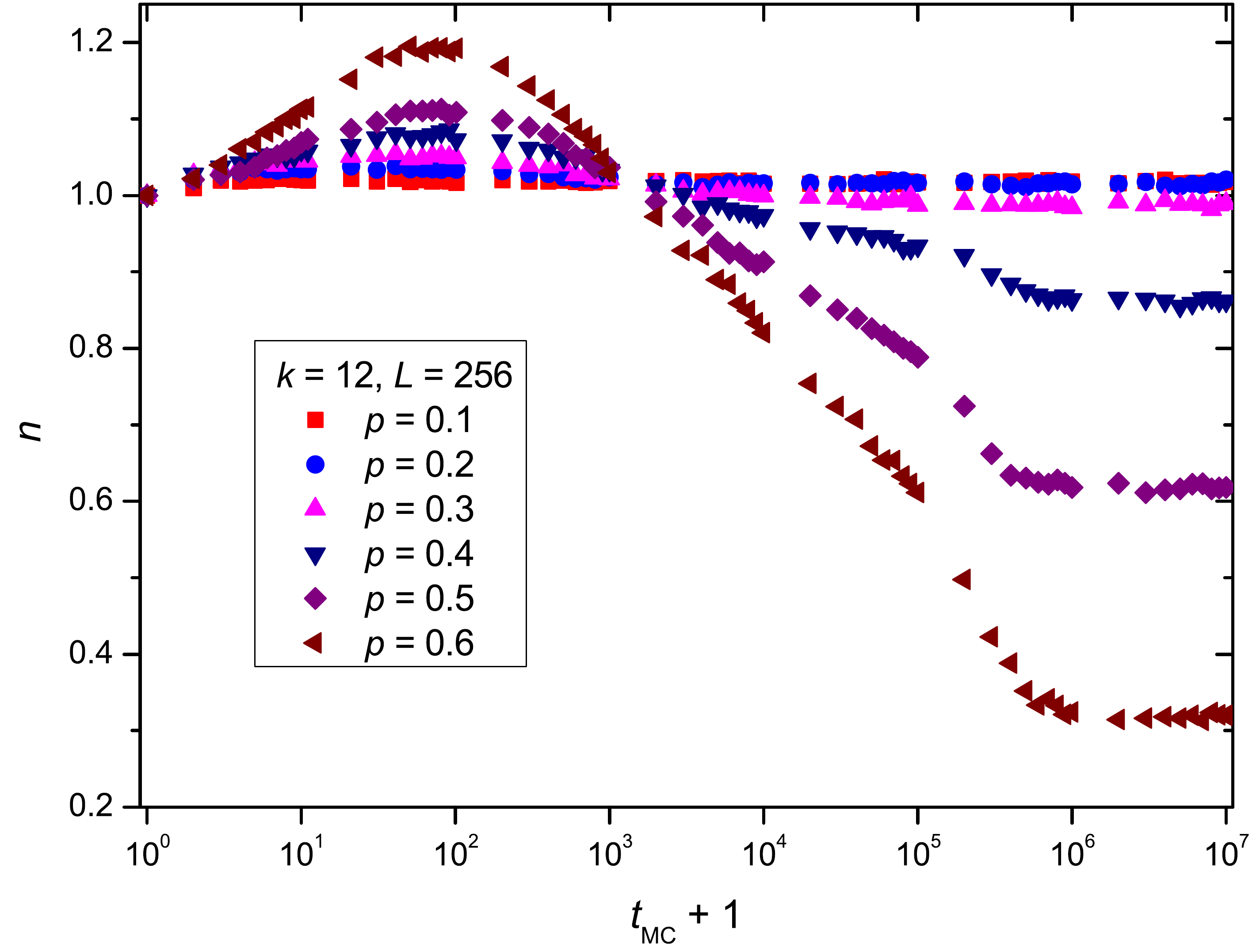}\hfill%
  (b)\includegraphics[width=0.45\linewidth]{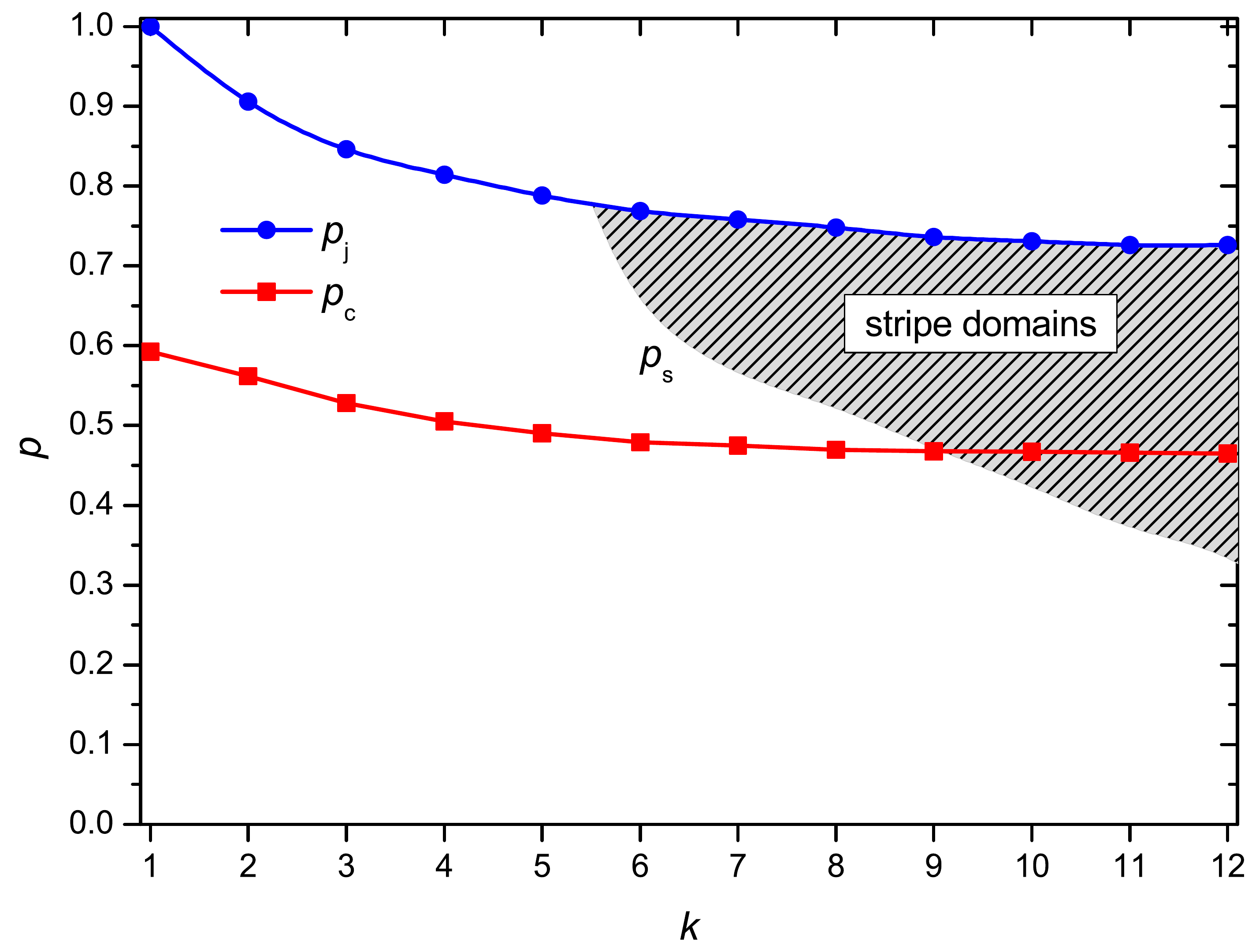}\\
  \caption{(a) Normalized number of clusters, $n$, vs MC steps, $t_{MC}$, for different concentrations of the $k$-mers, $L=256$, $k=12$, PBCs are applied.  (b) phase diagram in a $(k,p)$-plane; grey-shaded region corresponds to the values of $k$ and $p$ when stripe formation occurs; additionally, the percolation threshold and the jamming concentration are indicated for each value of $k$.}\label{fig:concentrations}
\end{figure}

A transition from a homogeneous steady-state to a patterned steady-state is continuous, i.e., the stripe domains become less pronounced when the initial concentration of $k$-mers decreases. At some initial concentration, the steady-state looks quite homogeneous. For instance, for the critical concentration, $p^\ast$, $0.3<p^\ast<0.4$ when $k=12$. For $p \geq 0.4$, the normalized number of clusters clearly demonstrates several stages of reorganization, while, for  $p \leq 0.3$, the curve looks similar to the curves for $k<6$, and for $p=p_j$ when stripe domain formation is absent (figure~\ref{fig:concentrations}).
Figure~\ref{fig:concentrations}b presents a phase diagram for the systems under study in a ($k,p$)-plane, when PBCs are applied. Here, the curves $p_c(k)$ and $p_j(k)$ show the dependencies of the initial percolation threshold and jamming concentrations at $t_{MC}=0$. The region with diagonal stripe formation in the steady state at $t_{MC} \gtrsim 10^6$ is shown in the phase diagram as the grey-shaded region. The diagonal stripes are not observed above the jamming concentration $p_j$ nor at small concentrations below some limiting value $p_s(k)$. The data show that at $k<6$ the initial jamming suppresses the diagonal stripe formation. However, the value of $p_s$ continuously decreases  with increased $k$ and at $k>9$; it becomes smaller than the percolation threshold $p_c$ for the given value of $k$. Previous data have evidenced that $p_c(k)$ dependencies go through a minimum at $k \approx 16$, but then the percolation threshold, $p_c$, increases at larger values of $k$~\cite{Tarasevich2012PRE}. The character of the $p_s(k)$ dependence at large values of $k$ is still unclear and requires further investigations.

\begin{figure}[htbp]
  \centering
  \includegraphics[width=0.5\linewidth]{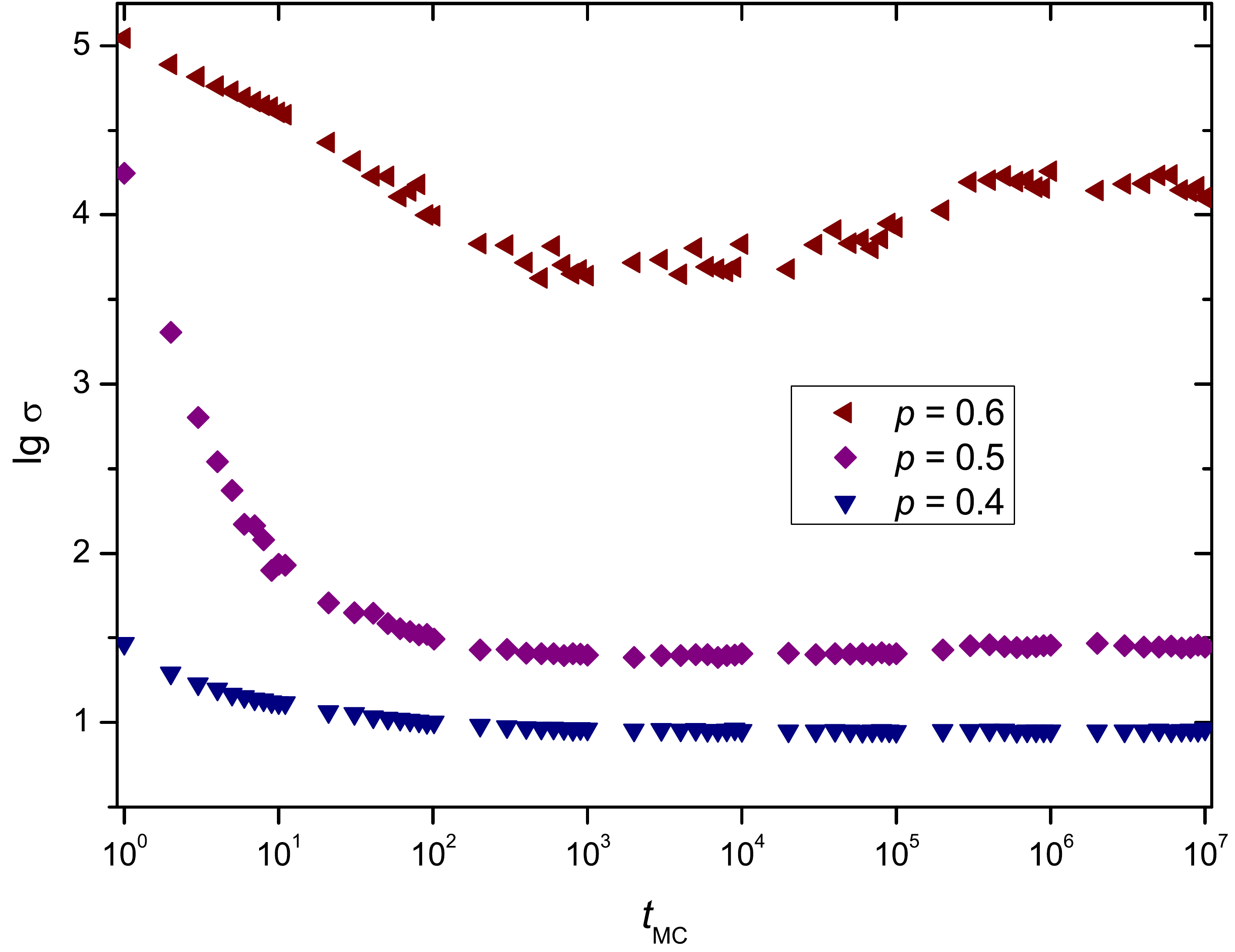}
  \caption{Examples of the temporal dynamics of electrical conductivity for $k=12$ and different values of concentration, $p$.}\label{fig:conductivityk12vsp}
\end{figure}

It is noticeable that pattern formation is accompanied by changes of electrical conductivity. When the concentration of particles slightly exceeds the percolation threshold, the system in its initial state, i.e., a disordered system, is a conductor.  Diffusional reorganization leads to a decrease in the electrical conductivity. Figure~\ref{fig:conductivityk12vsp} demonstrates the conductor--insulator phase transition for $k=12$ and $p=0.5$. This concentration is a little  greater than the percolation threshold, $p_c\approx 0.46$. In contrast, for $p=0.6$ (somewhat  greater than the percolation threshold) and $p = 0.4$ (below the percolation threshold), diffusion does not lead to a phase transition from conductor to insulator.

\subsection{Effect of anisotropy}\label{subsec:anisotropy}

\begin{figure}[htbp]
  \centering\small
  (a)\includegraphics[width=0.3\linewidth]{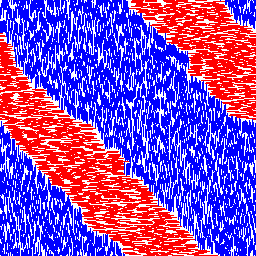}\hfill (b)\includegraphics[width=0.3\linewidth]{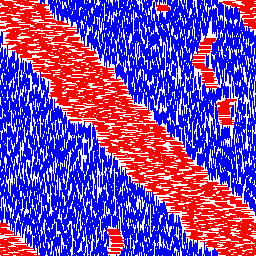}\hfill
  (c)\includegraphics[width=0.3\linewidth]{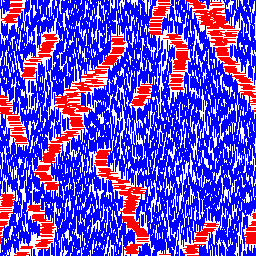}\\
  \caption{Examples of steady-state patterns for different values of the order parameter, $s$. (a)~$s=0.33$, stripe domains; (b)~$s=0.5$, stripe domains; (c)~$s=0.667$, wormlike domains.  $k=12$, $L=256$, $t_{MC}=10^7$, PBCs.}\label{fig:patternsanisotropy}
\end{figure}

In  two-species systems, stripe formation occurs not only in isotropic systems ($s=0$) but also in those systems with unequal  numbers of $k_x$-mers and of $k_y$-mers ($s \ne 0$). Thus, for $k=12$, the stripe domains occur up to $N_y/N_x = 3$ ($s=0.5$). For larger values of the ratio, no steady pattern forms until $t_{MC}=10^7$. Instead, a dynamic picture of moveable wormlike domains can be observed (figure~\ref{fig:patternsanisotropy}).

\subsection{Effect of boundary conditions}\label{subsec:bc}

When IBCs are applied, only two stages can be observed during the transient mode, viz., fluidization  and coarsening. The steady state demonstrates a spatial-temporal reorganization, i.e., each kind of $k$-mer tries to completely occupy one half of the lattice divided by a diagonal, e.g. the $k_x$-mers are located in the upper left corner while the $k_y$-mers are situated in the lower right corner (``yin-yang'' pattern). From time to time, regions built of $k_x$- and $k_y$-mers exchange their locations through irregular patterns (see video clip in Supplemental Materials).

\begin{figure}[htbp]
  \centering
  \includegraphics[width=\linewidth]{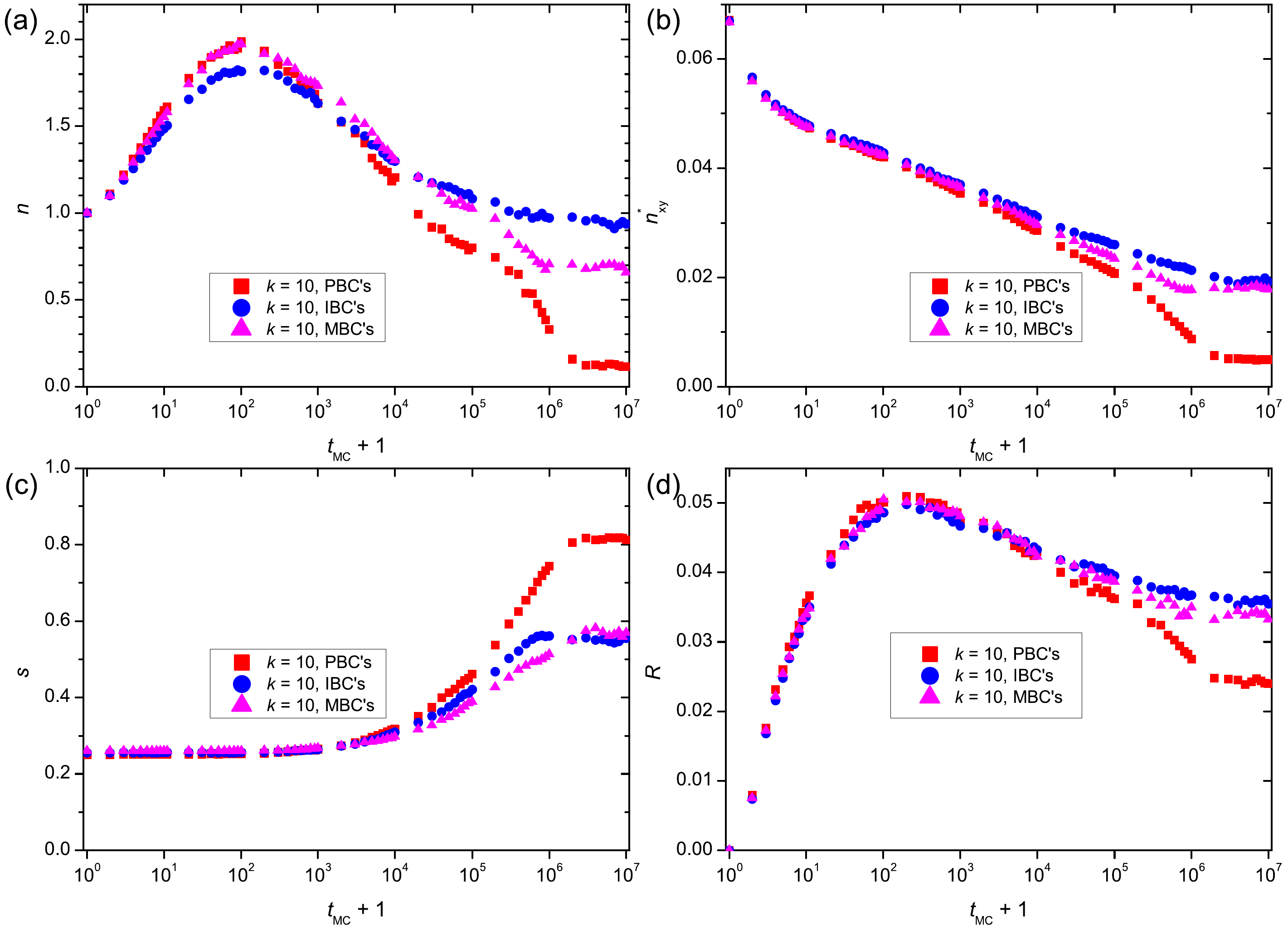}
  \caption{Comparison of behaviour of different quantities for different kinds of boundary conditions vs MC steps, $t_{MC}$, $k=10$, $L=256$.  (a)~Normalized number of clusters, $n$. (b)~Fraction of interspecific contacts, $n^\ast_{xy}$. (c)~Local anisotropy, $s$. (d)~Shift ratio, $R$.}\label{fig:AllBC}
\end{figure}

For $k<6$, all quantities vary with time in almost the same manner as in the case with PBCs. Nevertheless, for $k \geq 6$ one essential difference is observed, namely, a stepwise decrease in the quantities as $n$, $R$, and $n^\ast_{xy}$ are absent between $10^5 < t_{MC} < 10^6$ (figure~\ref{fig:AllBC}). Only the curves for the particular value of $k$ when stripe formation occurs are  presented. For the smaller values of $k$, no stripe formation is observed; changes in any of the quantities with time are insignificant for these values of $k$. The conductor--insulator phase transition does not occur. Although the local anisotropy increases with time, its final value is lower than in the case with PBCs (figure~\ref{fig:AllBC}).

For MBCs, when periodic boundary conditions are applied along one direction (horizontal) while  insulating boundary conditions are applied along the perpendicular direction (vertical) of the square lattice, after the two-stages of the transient mode, the system tends to form stripes. The stripes are unstable and change their spatial orientation (see video clip in Supplemental Materials).

Figure~\ref{fig:AllBC} compares the variations of the principal quantities with time. Here, the examples of curves for $k=10$ when stripe formation occurs are presented. For values of $k<6$, no stripe formation is observed; changes in $n$ with time are insignificant for these values of $k$. Only two stages of the transient mode are clearly seen. The local anisotropy, $s$, increases with time and this reflects the formation of the stripes. However, its final value is lower than in the case with PBCs (figure~\ref{fig:AllBC}(c)).

\subsection{Continuous model}\label{subsec:cont}
A simplest off-lattice 2D model of two-species diffusion with anisotropic nonlinear concentration-dependent diffusivities can be described by the set of equations
\begin{equation}
  \frac{\partial u }{\partial t}  = \frac{\partial}{\partial x} \left( D^s_1(u) \frac{\partial u}{\partial x}\right) + \frac{\partial}{\partial y} \left( D^s_2(u)   \frac{\partial u}{\partial y}  \right)  +  \frac{\partial}{\partial x} \left( D^c_1(v)  \frac{\partial v}{\partial x}\right)
  + \frac{\partial}{\partial y} \left( D^c_2(v)   \frac{\partial v}{\partial y}  \right), \label{eq:diffusionu}
  \end{equation}
  \begin{equation}
  \frac{\partial v }{\partial t}  =   \frac{\partial}{\partial x} \left( D^s_2(v)   \frac{\partial v}{\partial x} \right)
  + \frac{\partial}{\partial y} \left( D^s_1(v)   \frac{\partial v}{\partial y}  \right)
 +  \frac{\partial}{\partial x} \left( D^c_2(u)  \frac{\partial u}{\partial x} \right)
  + \frac{\partial}{\partial y} \left( D^c_1(u)   \frac{\partial u}{\partial y} \right),\label{eq:diffusionv}
\end{equation}
where $u$ and $v$ are the concentrations of the components, $D^s_i$ are the coefficients of self-diffusion, and $D^c_i$ are the coefficients of cross-diffusion. All diffusion coefficients are concentration-dependent. The first two terms on the right-hand sides of \eref{eq:diffusionu} and \eref{eq:diffusionv} correspond to self-diffusion, while the last two terms describe cross-diffusion.

Conceivably, the properties of this model might be similar to the properties of the lattice model described in section~\ref{sec:methods}. To examine the behaviour of the off-lattice model, we took the concentration-dependent diffusion coefficients corresponding to the case $k=12$ for the lattice model (figure~\ref{fig:DiffCoeff}).
$$
D_1^s(w) = 0.5 \exp(-0.09/w), \qquad   D_2^s(w) = 0.5 + 0.3 w + 1.6 w^2 - 3.2 w^3,
$$
$$
D_1^c(w) = 0.5 \exp(-0.18/w), \qquad D_2^c(w) = 0.5 \exp(-0.23/w),
$$
where $w$ is either $u$ or $v$.

Using the finite element method, we have solved the set of equations \eref{eq:diffusionu}, \eref{eq:diffusionv} in a square region with PBCs. For homogeneous distributions of the concentrations $u$ and $v$, the final concentration corresponded to the homogeneous distribution.

\section{Conclusion}\label{sec:conclusion}

We investigated a 2D lattice model of two-species diffusion. The species were considered as
hard-core (completely rigid) rodlike particles of two mutually perpendicular orientations.
We found that the diffusion coefficients of both self-diffusion and cross-diffusion are strongly nonlinear concentration-dependent. A non-monotonic concentration dependence of the coefficient of self-diffusion along the longitudinal axis of the $k$-mer is rather surprising.
We can speculate that such a feature reflects a  quasi-one-dimensional behaviour of the system.
 No qualitative changes in the coefficients of diffusion have been observed when $k$ increases from 5 to 6. Since $D^s_\parallel$ is much greater than $D^s_\perp$, $D^c_\parallel$, and $D^c_\perp$, any boundary between a region filled with $k_x$-mers and a region filled with $k_y$-mers tends to unbend, i.e. the curvature of the boundary decreases. At an interface between two regions filled with the different species, any  protruding $k$-mer retracts into a region with the same orientation of $k$-mers because movement in this direction is most probable. This fact can explain both cluster coarsening and the formation of stripes as well as the wormlike clusters. The larger a cluster, the smaller is the mean curvature of its boundary; the curvature of an ideal stripe boundary is equal to zero. Very small values of the coefficients of cross-diffusion at the jamming concentrations explain the presence of long-lived defects when the lattice sizes are large ($L=1024,2048$)~\cite{Lebovka2017PRE}. For a single $k$-mer, or a compact group the same species trapped inside a region of another species (e.g., $k_x$-mers inside a region of $k_y$-mers, or vice versa) their movement to an interface between the regions of the different species is restricted and can take a rather long time. Detailed investigation of effect of boundary condition on the behaviour of the system was performed for the lattice size $L=256$ and the jamming concentrations, $p=p_j$. We found that, for any boundary conditions under consideration, no oscillations or patterns were observed when $k < 6$. In contrast,  for $k\geq 6$, spatial or spatial-temporal pattern formation occurred, i.e., after a transient mode, the system came into a steady state with stable or periodically changed spatial patterns.

For those systems with equal number of species belonging to two different kinds, i.e., $k_x$-mers and $k_y$-mers, ($s=0$),
\begin{enumerate}
  \item for an initial homogeneous spatial distribution of $k$-mers, stripe domains form  even for $L/k \gtrsim 10^3$; the finite-size effect is important only for the last stages of pattern formation: although initial stages of reorganization are independent of the lattice size, the last stages, i.e., the stripe formation, are sensitive to the lattice size; nevertheless, steady state is independent on the lattice size;
  \item boundary conditions play a crucial role, namely, steady patterns in the form of stripe domains have been observed only for PBCs ;
  \item the self-organization is possible only in fairly dense systems;
  \item when the concentration of particles is slightly above the percolation threshold, an aging can be observed, i.e., a decrease in the electrical conductivity and the conductor--insulator phase transition at $t_{MC} \approx 10^2$.
\end{enumerate}
Moreover,  self-organization can occur in the systems with unequal numbers of $k_x$-mers and $k_y$-mers ($s \ne 0$).

An absence of pattern formation in the continuous model may be considered as indirect evidence that pattern formation in the lattice model is not the result of sole nonlinearity but also a hard-core (excluded volume) effect.
We can assume that some implicit correlations in the movements of the particles play an important role in the pattern formation. These correlations are more prominent when a system is dense and of moderate size, and the rodlike particles are long enough. We suppose that these correlations are a result of hard-core (excluded volume) interaction between the particles in the lattice model. These correlations are absent in the continuous model. This is a cause of the homogeneous final state in the continuous model. The discovery of such correlations is a particular problem.

The obvious shortcomings of our study are
\begin{enumerate}
  \item only relatively short particles ($k \leq 12$) were examined;
  \item only relatively small lattices (up to $L=2048$) were studied;
  \item a finite-size effect has been revealed, but this has not been studied in detail;
  \item long-time behaviour ($t_{MC} \gtrsim 10^8$) has not been explored.
\end{enumerate}
These shortcomings are a consequence of the fact that MC simulation is quite time-consuming even at a high-performance computer.

In future studies, one should keep in mind that, among the different quantities, the number of clusters isthe  most sensitive to changes of the internal structure of the system and, hence, it is a particularly useful quantity to use to classify different stages of the reorganization of the system.

\ack
The reported study was supported by the Ministry of Education and Science of the Russian Federation, Project No.~ 3.959.2017/4.6, and the National Academy of Sciences of Ukraine, Project No.~43/17-H.

\section*{References}
\providecommand{\newblock}{}

\end{document}